\documentclass[
aps,
12pt,
final,
notitlepage,
oneside,
onecolumn,
nofootinbib,
superscriptaddress,
noshowpacs,
centertags]
{revtex4}
\usepackage[english,russian]{babel}

\begin{document}
\selectlanguage{english}

\title{Optical Monitoring of the Central Region of the Globular Cluster M15
= NGC 7078: New Variable Stars}

\author{\firstname{A.P.} \surname{Zheleznyak}}
\email{zheleznyak@astron.kharkov.ua}
\affiliation{Research Institute of Astronomy, Kharkov National University,
Kharkov, Ukraine}

\author{\firstname{V.V.} \surname{Kravtsov}}
\email{scorpi@sai.msu.ru}

\affiliation{Sternberg Astronomical Institute, Universitetskii pr. 13,
Moscow, 119992 Russia}


\begin{abstract}
\centerline{ABSTRACT}
We obtained a series of more than two hundred $R$-band CCD images for the
crowded central ($115''\times77''$) region of the metal-poor globular
cluster M15 with an angular resolution of $0\farcs5-0\farcs9$ in most
images. Optimal image subtraction was used to identify variable stars.
Brightness variations were found in 83 stars, 55 of which were identified
with known cluster variables and the remaining 28 are candidates
for new variables. Two of them are most likely SX Phe variables.
The variability type of two more stars is uncertain. The remaining stars were
tentatively classified as RR Lyrae variables. A preliminary analysis
of published data and our results shows that the characteristics of RR Lyrae
variables in the densest part ($r<35''$) of the cluster probably change.
More specifically, the maximum of the period distribution of first- and
second-overtone (RR1, RR2) pulsating stars shifts toward shorter periods;
i.e., there is an increase in the fraction of stars pulsating with periods
$<0\fd3$ and a deficiency of stars with $0\fd35 \div0\fd40$. The ratio of
the number of these short-period RR Lyrae variables to the number of
fundamental-tone (RR0) pulsating variables changes appreciably. We found
and corrected the error of transforming the coordinates of variables
V128-155 in M 15 into the coordinate system used in the catalog of variable
stars in globular clusters.
\end{abstract}

\maketitle

\section{INTRODUCTION}

Globular clusters, particularly their densest central
regions, are among the objects whose observational
study, as well as the range and level of problems
to be solved, significantly depend on the limiting
angular resolution achievable in observations.
Obviously, new important results of the study of the
stellar composition and basic parameters of globular
clusters in our Galaxy and others have been obtained
from observations that have recently been performed
with the Hubble Space Telescope (HST) and ground-based
telescopes installed at sites with the best astronomical
climate.

The detection and study of photometrically variable
objects in crowded stellar fields belong to the
range of problems related to the investigation of the
stellar populations of globular clusters. Observationally,
variable stars of low (compared to RR Lyrae
stars) luminosity and small variability amplitude in
the central regions of these clusters are the most
difficult objects to detect. These primarily include
the stars that fall into the region of the so-called
blue stragglers in the color-magnitude diagram, as
well as the region of the turn-off point and near the
main sequence. They are represented by pulsating,
eclipsing, and cataclysmic variables. Our primary
objective was to attempt to detect such stars, along
with hitherto undetected RR Lyrae variables, in the
densest central part of the globular cluster M15
where their number can be large, considering the
parameters and the stage of dynamical evolution of
this cluster.

By its parameters, M15 is, in a sense, a unique object
among the Galactic globular clusters, especially
among those observable in the Northern Hemisphere.
According to the catalog by Harris (1996)\footnote{An updatable catalog
is accessible at http://www.physun. physics.mcmaster.ca/Globular.html.},
M15 is simultaneously among the clusters with the highest
mass, central star-crowding level, and density and is
at the evolutionary stage of postcore collapse. This is
so far the only Galactic globular cluster for which evidence
for the presence of a central intermediate-mass
black hole has been obtained (Gerssen et al. 2002,
2003). In addition, it is distinguished by a large population
(more than 150) of discovered variable stars (Clement et al. 2001)\footnote
{A full updatable catalog is accessible at http://www.astro.
utoronto.ca/people.html.}, the overwhelming majority
of which are RR Lyrae stars. Recently, this population
has been supplemented with new low-luminosity
variables (Jeon et al. 2001a, 2001b).

In Section 2, we describe the observational data, their reduction, and our
method of searching for variables and their photometry. Basic data on the
new variable stars are presented and described in Section 3. In Section 4,
we make a preliminary comparison of the parameters of the populations of RR
Lyrae variables in the central and outer parts of M15.

\section{OBSERVATIONS AND THEIR REDUCTION}

Our observational data form the basis for solving the problem formulated
above. These data aptly combine a high time resolution and a stable
subarcsecond angular resolution throughout the two sets (each three hours
long) of our optical monitoring of the central region in the globular
cluster M15.

\subsection{Observational Data and Preliminary Reduction}

We carried out our observations with the 1.5-m AZT-22 telescope at the
Maidanak Observatory (Mount Maidanak, Uzbekistan) during two consecutive
nights, on July 31 and August 1, 2001. The two nights were estimated to be
photometric. The detector was an ST7 CCD camera mounted at the short ({\it
f}/7.7)
Ritchey-Chretien focus of the AZT-22 telescope. The CCD array of the camera
has $765\times510$ light-sensitive pixels $9\times9$ $\mu$m in size, which
corresponds to a pixel angular size of $0.15''\times0.15''$ and a
$115''\times77''$ field of view of the camera for this AZT-22 configuration.
The observations were performed in a photometric band close to the standard
Johnson-Cousins $R$ band in a continuous frame-by-frame imaging mode with an
exposure of 1 min per frame. Given the image digitization and recording time,
the imaging rate was about 40 frames per hour. The upper culmination of M15
occurred at the middle of each set of observations. The total volume of our
observational data for the two nights was 248 frames (below, by the term
{\it frame}, we mean the two-dimensional array of counts that reproduces the
intensity distribution in the recorded {\it image}). The seeing estimated for
all of our frames from the full width of the seeing image at half maximum
(FWHM) showed a subarcsecond resolution almost for the entire volume of our
data. The average seeing for the two sets was $0\farcs74$; an appreciable
fraction of frames has a seeing FWHM $<0\farcs6$.

The preliminary CCD image reduction included the standard procedures of
dark-current subtraction, flat fielding, and cosmic-ray particle hit removal.
The mean dark current was estimated by averaging a series of 15 "dark"
frames, which gives a dark-current master frame. The normalized master frame
of a uniform flat field obtained by the median combination of a series of 15
frames containing the $R$-band images of twilight sky regions was used for
the flat fielding. Charged cosmic-ray particle hits in the form of
quasi-pointlike peaks were identified in each frame and removed by
interpolating the counts in the surrounding pixels. Previously, when
studying the photometric properties of the ST7 camera, we found a slight
nonlinearity of the output signal (about $10\%$ in the entire range), which
probably results from the adjustments of the ST7 output amplifier. Based
on tests, we obtained a dependence of the counts in the CCD image on the
intensity of the input signal varied in a known way and fitted it by a
polynomial. Using this fit, we corrected the intensity in each frame of
observational data for nonlinearity.

\subsection{The Data-Reduction Technique and Results}

The search for and photometry of variable stars in crowded stellar fields
are an important but, at the same time, difficult problem, particularly for
extremely dense objects like the centers of globular clusters. The
conventional approach to this problem realized in the standard software
packages for stellar-field photometry (DaoPhot, DoPhot) is based on the
point-spread function (PSF) fitting technique and involves the decomposition
or modeling of groups of close stars by using a particular PSF
representation method followed by the estimation of the coordinates and
magnitude for each of the stars in the group. Although most of the
stellar-field photometry results obtained to date are based on this approach,
we know several difficulties that arise when using the PSFphotometry methods.
At a high star density in the image, it becomes difficult both to estimate
the underlying sky background level and to properly estimate the parameters
of the PSF itself. Therefore, the problem of identifying and photometrically
measuring variable sources becomes much more complex.

The above difficulties stimulated the search for alternative approaches to
the problem of searching for variable objects in crowded stellar fields. In
particular, image-subtraction methods were developed. In general, the input
data in the search for variable objects are series of sequential frames that
contain the images of objects in a selected region of the sky. The idea of
the image-subtraction methods is to obtain information about the brightness
behavior of a source by analyzing the difference between the image in each of
the frames from the series and the image in a fixed reference frame. The
main problem in implementing the image-subtraction methods is the proper
reduction of the PSF in the reference frame to the PSF in the current frame.
The optimal image-subtraction (OIS) method that has recently been suggested
by Alard and Lupton (1998) elegantly solves this problem and allows a nearly
optimal difference between the images, i.e., limited by photon noise, to be
obtained. The OIS method is based on the following simple assumption: if two
images of the stellar field were resampled to a common coordinate system
(centered), then the counts in the overwhelming majority of the pixels will
be close if the PSFs of the two images are identical. The idea of the method
is to determine the convolution kernel \textit{Ker(u,v)} that reduces the
reference image $Im(x,y)$ to the current image $Im(x,y)$ by analyzing the
differences between the PSFs, i.e., to minimize the following difference in
all pixels:

\begin{equation}\label{1}
\sum_{x,y}\left([{{R}\otimes{Ker}}](x,y)-Im(x,y)-Bg(x,y)\right)^2\rightarrow\min
\end{equation}

where \textit{x,y} are the coordinates in the centered frame, $Bg(x,y)$
specifies the difference between the sky background counts, and the symbol
$\otimes$ denotes a discrete convolution. The latter is defined in a pixel
with \textit{x,y} coordinates as

\begin{equation}\label{2}
\sum_{u=x-p}^{x+p}\sum_{v=y-q}^{y+q} R(u,v)Ker(x-u,y-v)
\end{equation}

The convolution kernel \textit{Ker(u,v)} is specified on a separate
{[$2\cdot p+1$, $2\cdot q+1$]} array. Alard and Lupton (1998) suggested
representing \textit{Ker(u,v)} as a sum of fixed bivariate Gaussian
functions (the parameters $\sigma_k$) modifed by polynomials with a degree
not higher than $i,j$

\begin{equation}\label{3}
K(u,v)=\sum_{i,j,k,n} A_{n}e^{-\frac{(u^{2}+v^{2})}{2\cdot\sigma_k^2}}\cdot
 u^{i}\cdot v^{j}
\end{equation}

Representation (3) allows the minimization problem (1) to be linearized with
respect to the unknowns $A_{n}$. Thus, the parameters $A_{n}$ of the optimal
kernel \textit{Ker(u,v)} can be determined by solving the system of linear
equations by the least squares method. Without dwelling on the detailed
description of the method by Alard and Lupton (1998), we note important
advantages of the OIS method. When the system of normal equations is
constructed, all pixels, i.e., in principle, complete information about the
difference between the PSFs of the two images, are used to find the
parameters $A_{n}$. The OIS method works more reliably when the star density
in the image increases, because more pixels contain information about the PSF
difference in denser fields. The intensity-conservation condition can be
easily deduced by appropriately normalizing the kernel representation
parameters $A_{n}$. Thus, any variations in transparency, exposure time, and
the like between the images are automatically eliminated. Finally, the PSF
variations over the field can be taken into account in the solution by
introducing a dependence of the coefficients $A_{n}$ on the coordinates.
In the case of proper subtraction, only the variable part of the intensity
remains in the difference image. Thus, only variable objects that are
unaffected by the surrounding field can be detected and measured. The
difference images can be photometrically measured in a standard way-using
aperture or PSF photometry. It should be noted that difference-image
photometry gives the variable part of the flux from the object; i.e., when
applied to the time sequence of images, it gives a light curve without the
constant component. If the total flux is needed for a particular task, then
the light curve should be calibrated by performing the photometry of the
corresponding objects in the reference image by a standard method. The idea
of the OIS method underlies the software that we developed to process the
images of M15. The reduction algorithm consisted of the following steps:

(1) \emph{Reference-image synthesis}. We selected nine frames with the best
seeing (FWHM$\sim{0\farcs55\div0\farcs58}$) from the data of the two sets.
Using a system of selected reference stars, we determined the image
shifts/rotations in each of the selected frames relative to the coordinate
system associated with the CCD array. Subsequently, all of the images were
resampled to the same coordinate system by using bivariate spline
interpolation and were combined pixel by pixel with median weighting. The
resulting reference image has FWHM=$0\farcs56$ and is virtually free from
local defects.

(2) \emph{Image centering}. For the subsequent reduction, all of the images
for M15 from the frames of the two observing sets were centered on the
reference image. A correlation algorithm was used to determine the
shift/rotation parameters. After the interpolation to the coordinate system
of the reference image, the new sequence of centered frames was saved as a
series of files.

(3) \emph{Image subtraction}. The main procedure in calculating the
difference between the current and reference images involves determining
the parameters $A_{n}$ of the optimal convolution kernel (3), which reduces
the reference image to the current image. We specified the convolution
kernel \textit{Ker(u,v)} as a sum of three bivariate Gaussian functions
with fixed parameters {$\sigma_k$ 1.3, 2.25, and 3.9} in combination with
bivariate modifying polynomials of degrees 4, 3, and 2, respectively. The
array size for \textit{Ker(u,v)} was $31\times31$ pixels; the difference
between
the sky background counts $Bg(x,y)$ was specified by a bivariate polynomial
of the first degree with appropriate coefficients. In our calculations, we
broke down the frames of the reference and current images into nine equal
rectangular fragments and determined the parameters $A_{n}$ of the optimal
kernel \textit{Ker(u,v)} in each of these fragments. The coordinate
dependence of the PSF may be disregarded within each fragment, which
decreases the number of sought-for parameters to 36 and significantly
reduces the expenditure of time on the calculations.

Having calculated $A_{n}$ for a given fragment of the current frame, we
constructed an optimal kernel \textit{Ker(u,v)}, convolved the corresponding
fragment of the reference imagewith this kernel, and subtracted the
convolution result from the current fragment. Analysis of the resulting
difference image makes it possible to find and fix the coordinates of the
pixels that significantly (by more than $3\sigma$) deviate from the mean in
a given region of the difference image and that are associated with variable
objects or local defects. Using the coordinates of the marked pixels, we
eliminated them from the original system of equations and obtained an
improved solution by repeating the above procedure of calculating the
parameters $A_{n}$.

Having performed the calculations in each of the nine fragments of the
current frame, we combined the difference images and formed a full frame of
the difference image, which was saved as a file. The absence of any steps at
the fragment boundaries in the full difference image is an indicator of
proper subtraction.

Figure 1 illustrates how the described algorithm works.
The original fragments of the (Fig.~1a) reference and (Fig.~1b) current
images for M15 are shown in the upper part of the figure. Although the PSF
in the fragment of the current image (Fig.~1b) is appreciably broader and
has a complex elongated shape, the OIS method allows us to find a solution
for the optimal convolution kernel \textit{Ker(u,v)} (Fig.~1c) and perform
the subtraction; the corresponding fragment of the difference image is
shown in Fig.~1d. The variable stars ZK6 and 7 (the numbers are from our
list; the prefix "ZK" means that the star is a candidate for new variables)
are clearly identified; we see that in the current fragment, star ZK6 is
fainter, while variable 7 is brighter than they are in the reference image.
The modulus of the intensity distribution in the difference images of these
variables, naturally, corresponds to the PSF of the current image. We also
see from Fig.~1d that the subtraction of the surrounding nonvariable stars,
including those overlapping with variables ZK6 and 7, was performed properly.
The appreciable residual-intensity .uctuations in the difference image
(Fig.~1d), which are mainly attributable to photon noise, are observed only
at the positions of the brightest stars (see, e.g., to the left and below
variable 7).

Having reduced the data from the two sets of observations of M15 using the
described algorithm, we obtained the corresponding series of difference
image frames for the subsequent analysis. The procedure for finding
variable objects can be based both on the analysis of the behavior of each
pixel with time, i.e., within the derived sequence, and on the analysis of
the average frame from several sequential difference-image frames. We used
both approaches to find variable objects. As a result of the selection, we
identified a group of pixels with appreciable variability and determined
their coordinates (the centroid of the modulus of the difference intensity
distribution) in the coordinate system of the reference image. A list of
potentially variable objects with the coordinates of their centroids was
saved in a separate file for subsequent reduction. The formal measurement
error of the centroid coordinates for most objects was about 0.1 pixel
($\sim0\farcs02$). The ultimate selection of candidates for variable stars
was based on photometric data.

We used aperture photometry, which is easier to realize, to measure objects
in the difference image. An aperture 10 pixels (1\farcs5) in diameter was
used to measure the intensity in the difference image. Based on the centroid
coordinates from our list, we measured the intensity of each residual from
all of the difference images. Objects with a deviation from the mean of more
than 3\% within the entire series were classified as candidates for variable
stars. Since all of the images, both the original ones before the
subtraction and the difference ones, were reduced to the same coordinate
system, it will suffice to formally estimate the photometry error from the
total-flux fluctuations within the measuring aperture. Placing the measuring
aperture in star-centroid coordinates into the frame of the original
(before the subtraction) image and summing the intensities, we estimate the
rms measurement error $\sigma_m$ from the total intensity in the aperture
(under the assumption of Poisson statistics):

\begin{equation}\label{4}
\sigma_m=\frac{1}{\sqrt{\sum_{i}{\left(I_{0i}*g\right)}}}
\end{equation}

The summation is over the pixels within the aperture
(index $i$), $I_{0i}$ is the intensity in pixel $i$ before the subtraction
(including the sky background), and $g$ is the number of electrons per ADC
unit. Our estimates of the formal error $\sigma_m$ vary within the range
$0\fm03-0\fm005$, depending on the object's brightness.

Since we have a sample of almost homogeneous measurements at our disposal,
we can also directly estimate the photometry error by analyzing the scatter
of individual points about the mean (e.g., on monotonic segments of the light
curve after the elimination of the trend). The formal and direct estimates of
the photometry errors made for several objects are in satisfactory agreement.
Therefore, we considered it possible to use the formal estimate for the
remaining objects.

Our reduction of the data from the two sets of observations in a
$115''\times77''$ central region of M15 revealed 83 stars with appreciable
variability on short time scales. The coordinates and light curves relative
to the reference frame form the basis for the subsequent analysis of our
sample of candidates for variable stars.

\section{NEW VARIABLE STARS}

\subsection{The Identification of Variables and the Determination
of Their Coordinates}

The overwhelming majority of the new variables in the densest central
($r<20''$) region of M15 were discovered in the past decade by Ferraro
and Paresce (1993) and Butler et al. (1998), who used observations with the
HST and the William Herschel (4.2-m) telescope, respectively. A series of
several tens of cluster images with a subarcsecond angular resolution
allowed the latter authors to confirm the variability of 15 stars from the
list by Ferraro and Paresce (1993), to discover 13 new variables (apart
from the two known variables V83 and V85), to obtain the phased light curves,
and to determine the variability periods for all of these stars. In the
catalog by Clement et al. (2001), these 28 new variables are designated as
V128-155. However, their coordinates in this catalog proved to be erroneous
in the sense that they do not correspond to its coordinate system. This
fact was established during the identification by using the cataloged and
original data. The formulas in the catalog that were used to recalculate the
coordinates are erroneous.

For the above stars, Butler et al. (1998) provided both highly accurate
rectangular coordinates (in pixels) and a finding chart. Therefore, we were
able to reliably identify 19 of the 28 stars with the corresponding stars
from our list: V128-142 (46, 61, 58, 59, 60, 57, 56, 53, 51, 50, 38, 42, 35,
30, 36), V144 (25), V145 (24), V152 (45), V155 (66). The numbers of the
variables from our list are given in parentheses. It does not include the
nine remaining stars, because they exhibited no light variations in our
images. Two more stars (565 and 1417) from the list by Butler et al. (1998),
which most likely correspond to V83 and V85, are also reliably identified
with stars 21 and 17 from our list, respectively.

Apart from the stars considered above, several dozen of other known variables
from the catalog by Clement et al. (2001), which is known to be an extension
of the catalog by Sawyer Hogg (1973), fall within our observed field of M15.
Unfortunately, the main drawbacks related to the rectangular ($x'', y''$)
coordinate system of the catalog by Sawyer Hogg (1973) passed to the catalog
by Clement et al. (2001). For well-known reasons, the accuracy of these
coordinates in the catalog is low for most of the variable stars. In
particular, as Kadla et al. (1988) pointed out, the positions of variables
in M15 were determined in two rectangular coordinate systems with a
difference between the zero points larger than $2''$ along each of the axes
back in the first quarter of the past century. Therefore, in some cases,
confusion with the identification of variables arises, because the
differences between their positions determined by different authors reach
several arcseconds. In the absence of finding charts, this circumstance
makes it difficult to unambiguously identify new variables with previously
discovered ones. This is especially true for the central parts of
variable-rich globular clusters, such as M15, where the probability of a
close neighborhood of variables is high. We also ran into this problem.
However, Kadla et al. (1988) published a finding chart for variable stars
in the part of the cluster concerned. Therefore, we were able to solve the
problem more or less reliably with a number of controversial cases
(see below) that arose when we used the cataloged information about the
coordinates of the variables. The authors of the paper mentioned above
reduced the coordinates for all of the variables known at that time
in M15 to the same coordinate system and pointed out the cases where the
same stars were denoted in the catalog by different numbers.

The final identification procedure was performed as follows. We determined
the variables from the catalog by Clement et al. (2001) that were most
reliably identified with stars from our list (mostly in the outer parts of the
observed cluster field) and transformed the coordinate system of the former
to the coordinate system of the latter. Subsequently, we determined the
equatorial coordinates for the stars from the list by Butler et al. (1998),
our list, and the catalog by Clement et al. (2001), naturally, by excluding
variables V128-155 from it. In this case, reference stars from the catalog
of M15 stars by Yanny et al. (1994) were used for referencing to the
equatorial coordinate system. For the stars from all of the lists, we
determined the differences $\Delta\alpha$ and $\Delta\delta$ from their
equatorial coordinates relative to the M15 center (the position of the
object AC 211), whose coordinates, $\alpha_{2000}=21^h29^m58\fs26$ and
$\delta_{2000}=+12\degr10\arcmin02\farcs90$, were taken from the same paper.
Taking into account the small size of the cluster field under consideration
and its relatively small angular distance from the celestial equator, we
simultaneously take these differences as the rectangular coordinates
($x''$ and $y''$) in the system of the catalog by Clement et al. (2001).

The identification result is shown in Fig.~2. This figure clearly
illustrates the above-described reliable and unambiguous identification of
a total of 21 stars from our list with the corresponding stars from the
list by Butler et al. (1998). The situation with the identification of
stars from the catalog by Clement et al. (2001) proved to be less certain
and reliable because of several controversial cases. As we see from Fig.~2,
each of the stars with numbers 2, 8, 15, 19, 20, 26, 27, 28, 33, 73, and 75
clearly do not coincide with any of the nearby catalogued stars. However,
on the finding chart from Kadla et al. (1988) mentioned above, they are
identified with the catalogued variables V47, V120, V119, V117, V84, V109,
V100, V115, V95, V106, and V107, respectively. For all of the remaining
identified stars, the difference between their coordinates ($\Delta x''$ and
$\Delta y''$) and those of the catalogued variables proved to be much
smaller and such that $r<2''$, where $r=\sqrt{(\Delta x'')^2 + (\Delta
y'')^2}$. Below, we list the catalogued stars that (apart from the above
eleven stars) were identified with stars from our list (their corresponding
numbers are given in parentheses): V33 (82), V56 (1), V64 (83), V68-71
(76, 81, 77, 79), V73 (54), V75-77 (43, 49, 66), V79 (16), V81-83 (71, 70,
21), V85-90 (17, 29, 12, 41, 72, 7), and V92-94 (31, 9, 40). In addition,
the position of variable 48 in the cluster closely ($<0\farcs1$) matches
the position of the well-known object AC 211 (Auriere and Cordoni 1981).
Auriere et al. (1984) identified this object with the X-ray source X2127+119
located at the center of M15.

We identified a total of 55 of the 83 stars from our list with known
variables, while the remaining 28 stars were classified as candidates for
new variables in the globular cluster M15. Their numbers are given in
Table 1. This table also presents the coordinates of these stars, which,
as was noted above, are the differences $\Delta\alpha''$ and
$\Delta\delta''$ relative to the coordinates of the cluster center in
arcseconds. To identify new variables (among all of the 83 stars that we
discovered), in describing them in the text, the table, and the figures, we
added the prefix "ZK" to their numbers.

Table 2 gives corrected coordinates for V128-155. These coordinates are also
the differences $\Delta\alpha''$ and $\Delta\delta''$, which may be taken
with a sufficient accuracy as the rectangular coordinates ($x'', y''$)
relative to the cluster center in the system of the catalog by Clement et al.
(2001).

\subsection{The Classification of New Variables}

The basic characteristics of our observations and the achieved accuracy of
the (relative) photometry allowed us to reliably establish the fact of light
variations in the candidates for variable stars. Unfortunately, however, the
total duration of our optical monitoring was not long enough for the same
reliable and definite classification of a large number of them. We primarily
attribute this to the determination of the subtype of RR Lyrae variables,
considering the highest probability of their presence among the discovered
new variable stars. At the same time, we do not rule out the erroneous
assignment of variables of other types whose periods exceed the duration of
each of the two monitoring intervals to the latter. Nevertheless, our
quasi-continuous fragments of the light curves make it possible to
{\it tentatively} estimate the type of the discovered variables from the
shape and characteristic features of these curves, the range of brightness
variations, etc. This estimate is presented in the column "Type" of Table 1.
In addition, this table gives data on the detected brightness variation
$\Delta R$, which is not an amplitude value for some of the variables.
In designating the different subtypes of RR Lyrae stars, we followed the
new system of designations for these subtypes adopted in the catalog by
Clement et al. (2001): RR0 corresponds to fundamental-mode pulsations;
RR01 corresponds to a double pulsation mode (in the fundamental tone and
the first overtone); and RR1 and RR2 correspond to first- and second-
overtone pulsations, respectively.

Although the mean positions of the new variables in the color-magnitude
diagram for M15 cannot be determined from our data, we managed to identify
most of them with stars in the catalog from van der Marel et al. (2002).
It contains photometric data for almost 32 000 stars in the central region
of M15 that was observed with the HST. Therefore, we determined the
positions of the following identified new variables in the $V-(B-V)$
diagram constructed from the data of this catalog (Fig.~3): ZK3
(21552), ZK4 (20097), ZK5 (20118), ZK6 (11423), ZK10 (11776), ZK11 (24953),
ZK13 (26619), ZK14 (13564), ZK18 (25693), ZK22 (1855), ZK23 (505), ZK34
(27077), ZK37 (3393), ZK39 (8568), ZK44 (2997), ZK47 (5768), ZK52 (10356),
ZK55 (2321), ZK62 (10041), ZK64 (4894), and ZK68 (10344). The new variables
ZK34 (876), ZK37 (793), ZK39 (784), ZK44 (722), ZK47 (699), ZK52 (511),
ZK64 (332), ZK67 (267), ZK68 (210), and ZK80 (46) were also identified with
stars of the photometric catalog from Yanny et al. (1994), which was
compiled from early HST observations. The star numbers from these catalogs
are given in parentheses. The identified variables are indicated by circles
in Fig.~3. With the exception of five stars (for more detail, see below),
the positions of the remaining stars in the diagram are in good agreement
with their classification as RR Lyrae variables.

As we see from Table 1, we managed to discover two stars, ZK62 and ZK68,
that are, undoubtedly, not RR Lyrae variables. The amplitude and period
($<0\fd08$) of their brightness variations allow us to classify these stars
with a high probability as SX Phe variables, especially since, as we see
from Fig.~3, ZK62 and ZK68 fall into the region of blue stragglers. Details
on ZK62 and ZK68 will be presented in a separate publication. Note that,
currently, only one variable of this type is known in M15. Recently, Jeon
et al. (2001a) detected it at a distance of several arcminutes from the
cluster center. Thus, given the results of our study, the number of SX Phe
stars discovered in the globular cluster M15 reached three. In contrast to
ZK62 and ZK68, we failed to determine the variability type of the other
two variables, ZK32 and ZK47. As follows from the photometric data of van
der Marel et al. (2002), the latter proved to be among the red-giant-branch
(RGB) stars above the horizontal branch in the $V-(B-V)$ diagram, while
according to the data of Yanny et al. (1994), it is located in the $V-(V-I)$
diagram above (by more than $0\fm5$) the blue part of the horizontal branch.
ZK10 and ZK39 also proved to be among the RGB stars in Fig.~3. However, the
latter, according to the data of Yanny et al. (1994), is located near the
horizontal branch, while in Fig.~3 it lies below this branch by more than
$1\fm0$. Therefore, it may well be that we identified it erroneously in the
catalog by van der Marel et al. (2002), because the difference in
coordinates between it and the star identified in the catalog was larger
than that for other variables.

Note, in addition, that star 48, i.e., the object AC 211, in our images
exhibited complex brightness variations. On the first night of our
observations, they appeared periodic, with $\Delta R \approx 0\fm1$ and a
period on the order of $0\fd1$, while on the second night these variations
manifested themselves in the relatively slow brightness increase by $\Delta
R \approx 0\fm2$, only with a hint at faster brightness oscillations (at
the photometry error level).

Our preliminary classification by types of new variables is illustrated by
examples of light curves for twelve of these stars in Fig.~4. Figures 4a,
4b, and 4d show the light curves of RR0, RR1, and RR2 stars, respectively.
The light curves of the stars mentioned above whose variability type could
not be determined are shown in Fig.~4c. This figure also shows the light
curve for the object AC 211. The star numbers are indicated near the
corresponding curves. For convenience, we arbitrarily displaced the light
curves along the vertical axis ($\Delta R$). Formally, there are no RR01
stars among the candidates for new variables. However, this is not an actual
fact. It will be possible to reliably establish whether particular stars are
of the RR01 subtype only after the determination of their periods. As we see
from the figure, we conditionally classified the stars in which the recorded
brightness variation allows the amplitude to be determined and, at the same
time, $\Delta R$ is generally within $0\fm25$ and the shape of the light
curve is most likely nearly sinusoidal as being of the RR2 subtype. In this
case, there is reason to believe that the second interval of our observations
(its duration is $0\fd128$ compared to the duration $0\fd107$ of the first
interval of our observations) spans about half the pulsation period of the
RR2 stars under discussion; i.e., their periods are $<0\fd3$. The stars
classified as RR1 exhibit brightness (including amplitude) variations larger
than $0\fm25$ and an indistinct asymmetry in the light curves. For some of
them, the pulsation periods can also be $<0\fd3$, as those for the stars
classified as RR2. The detected brightness variations in RR0 stars proved
to be, on average, even larger than those in RR1 stars. In addition,
particular characteristic features, such as an asymmetry, a sharp peak, etc.,
showed up in their light curves. Nevertheless, in several cases, the
determination of the subtypes of new variables is controversial or
conditional.

\section{COMPARATIVE CHARACTERISTICS OF VARIABLE STARS IN THE CENTRAL AND
OUTER PARTS OF M15}

Almost ten years ago, Stetson (1994) performed multicolor photometry of
stars in the central part of M15 by using a large number of CCD images with
a subarcsecond angular resolution and an improved procedure of stellar
photometry. He found several important changes in the stellar composition
that were observed in the cluster region at $r<30''$, especially in its
densest part ($r<12''$), and showed up in the color-magnitude diagram. In
particular, these changes manifested themselves in a significant growth of
the population of stars that fell into the region in the diagram between the
turnoff point and the horizontal branch, as well as in an apparent
reddening of the horizontal branch itself. It would be quite natural to
assume that the effects resulting in the detected changes in the
color-magnitude diagram could also be reflected on the population of
variables located in the same part of the cluster and falling into the same
region of the diagram. One of the possible changes is quite a natural and
expected result of the dynamical effects in the central part of M15 or more
specifically, an increase in the number of variables associated with binary
stars and blue stragglers. This is confirmed by the study of the central
parts of post-core-collapse globular clusters. In particular, in NGC 6397,
Kaluzny and Thompson (2002) discovered several variables, among which are
eclipsing, cataclysmic, and SX Phe stars. We also managed to discover two SX
Phe variables. However, taking into account one of the above results of
Stetson (1994), one might expect the actual number of such stars in the
central part of M15 to be significant. Whether any changes could affect the
population of RR Lyrae variables proper is quite a different matter.

A preliminary (including frequency) analysis of the light curves indicates
that there may be a significant fraction of stars with periods $<0\fd3$
among the new variables classified as RR1 and RR2 and located in the cluster
region at $r<35''$. The most realistic estimation gives a lower limit on the
order of seven (of 15), i.e., more than 40\%. We noticed that only five
(less than 20\%) among the stars of the above subtypes in the catalog by
Clement et al. (2001) with measured periods and located at distances
$r>35''$ from the center of M15 (i.e., with the exception of V128-155, V83,
and V85) satisfy this condition; none of them has a period $<0\fd27$.
However, the fraction of such stars (more than 60\%) significantly increases
among the variables of the above subtypes located in the central region of
M15 (i.e., among V128-155, V83, and V85). Almost half of them have periods
$<0\fd27$. The change in the ratio of the numbers of RR Lyrae stars pulsating
with periods $<0\fd3$ and those pulsating in the fundamental tone (RR0)
proves to be more significant: 80\% in the central region compared to
15\% outside it. Yet another significant difference is that only two
(of the 25 sample stars) among the variables in the central part of the
cluster have periods in the range $0\fd35 \div0\fd40$, while there are
almost a third of such variables of their total number, more specifically,
24 of the 76 sample stars, in the outer parts of the cluster. A similar
change probably also pertains to the numerous population of RR01 stars in
the outer parts of the cluster.

Based on the data from Butler et al. (1998) and from the catalog by Clement
et al. (2001) analyzed above, we constructed the histograms (Fig.~5) that
clearly show the described differences between the period distributions of
the RR Lyrae variables located in different parts of M15. The number of stars
in the outer regions of the cluster allows them be separated into three
subpopulations (equal in number to the population in the central part)
located in annular zones at different distances from the center of
M15: 26 stars with $r>175''$ and 25 stars each with $90''<r<175''$ and
$35''<r<90''$. This separation makes it possible not only to analyze the
period distributions of variables at different distances from the
 center
of M15 in its outer regions but also to compare each of them with the
distribution of stars in the central region. The number of stars in each
sample is the same. The derived histograms of the corresponding
distributions are represented by dotted lines in Fig.~6 for stars at
different distances from the center of M15. As in Fig.~5, the solid line
indicates the period distribution of the variables in the central region.
We see from a comparison of the histograms that there is no clear evidence
of any systematic differences between the period distributions of the RR
Lyrae variables in the three outer parts of M15 and that each of them
significantly differs from the period distribution in the central part.
This suggests that the changes (if they are real) occur almost abruptly
somewhere at $r<35''$, where the star density significantly increases and
where Stetson (1994) found apparent changes in the color-magnitude diagram
of the cluster.

Undoubtedly, the size of the sample of variables in the central region of
the cluster (25 stars) is much smaller than the size of their sample
(76 stars) in the remaining part of M15 and is not yet sufficient
to draw ultimate and reliable conclusions. However, the presented
differences are too striking to be left unnoticed. In addition, a
preliminary analysis of our observations of the new variables reveals the
same tendency. Of course, we cannot rule out the possibility that the
differences being discussed are attributable to selection effects. One of
these effects may stem from the fact that, historically, the variables in
the outer parts of M15 were studied mostly by using the methods of
photographic photometry of these stars. In contrast, the data on the
variables in the central part of the cluster were obtained mainly in the
past decade by using much more sophisticated and efficient image recording
and reduction techniques. This imbalance between the possibilities of
observational studies could lead to the fact that some of the low-amplitude
(and, hence, on average, shorter-period) RR Lyrae variables in the outer
parts of M15 have proven to be simply undetectable so far. At first glance,
this assumption seems quite justified. However, it should be borne in mind
that, while studying several tens of known RR Lyrae stars in a wide cluster
field using a CCD array, Silbermann and Smith (1995) discovered only one
such low-amplitude variable of this type, namely, V113.

With the addition of data on the periods, light curves, and positions in the
color-magnitude diagram of our discovered stars and the already known but as
yet unstudied variables in the central part of the cluster, the sizes of the
samples of variables located in the outer and central regions of M15 will be
comparable and quite sufficient for a more substantive analysis.

\section{CONCLUSIONS}

We carried out two sets of optical monitoring of the central region in the
globular cluster M15 with a subarcsecond angular resolution and a total
duration of about six hours using the 1.5-m telescope. As a result, we
obtained more than two hundred $R$-band images of the cluster. The reduction
of our data using the optimal image-subtraction method of Alard and Lupton
(1998) revealed brightness variations in 83 stars. Twenty eight of them are
candidates for new variables, which constitute the largest population of
new variables discovered in one research work in the past 50 years of the
study of variable stars in M15. Apart from the two stars whose variability
type could not be determined, the other two stars are likely to be SX Phe
variables, while the remaining stars were tentatively classified as RR
Lyrae variables. Published data on the variables of this type located in the
central region of the globular cluster and a preliminary analysis of our
results show that in the densest part ($r<35''$) of the cluster, the
maximum of the period distribution for first- and second-overtone pulsating
(RR1 and RR2) stars probably shifts toward shorter periods. In addition to
an increase in the fraction of these stars pulsating with periods $<0\fd3$,
there is a deficiency of stars in the range of periods $0\fd35 \div0\fd40$
compared
to the period distribution for the population of variables in the farther
outer parts of M15. The ratio of the number of variables with periods
$<0\fd3$ to the number of variables pulsating in the fundamental tone (RR0)
also changes. We found and corrected the error of transforming the
coordinates of variables V128-155 to the coordinate system of the catalog
by Clement et al. (2001).

\acknowledgments
We are grateful to the referee, N.N. Samus', for helpful remarks.

\
\
{\bf Translated by V. Astakhov}

\begin{figure}
\setcaptionmargin{5mm}
\onelinecaptionsfalse
\includegraphics[width=14.0cm]{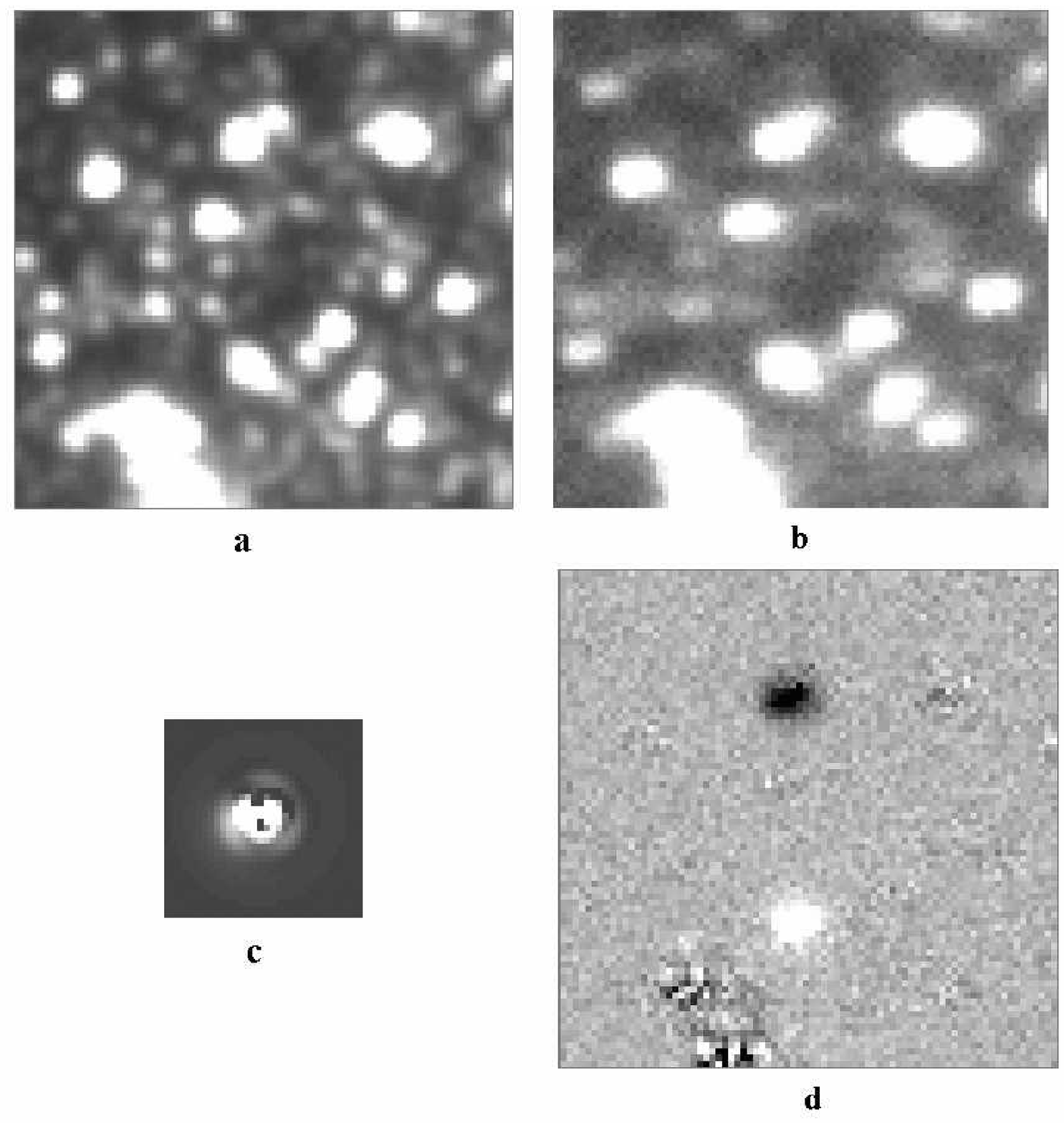}
\captionstyle{normal}
\caption{An illustration of how the optimal
image-subtraction (OIS) algorithm works. The original fragments of the (a)
reference and (b) current images for M15; the fragment size is
$12''\times12''$; (c) the relative-intensity
distribution in the optimal kernel \textit{Ker(u,v)} calculated from the
reference and current images by using the algorithm described in the text.
The sizes of the shown image for \textit{Ker(u,v)} are $31\times31$
independent pixels; the gradations for the visualization were chosen to
optimally reproduce the faint outer wings. The relative-intensity
distribution in the image of the optimal convolution kernel reflects the
complex PSF shape for the current image (b); (d) the result of the
subtraction of the fragment of the reference image (a) convolved with the
optimal kernel from the fragment of the current image (b). The gradations
were chosen to represent both positive (lighter) and negative (darker)
fluctuations of the residual intensity near a zero mean (gray).}
\end{figure}

\begin{figure}
\setcaptionmargin{5mm}
\onelinecaptionsfalse
\includegraphics[width=16.5cm]{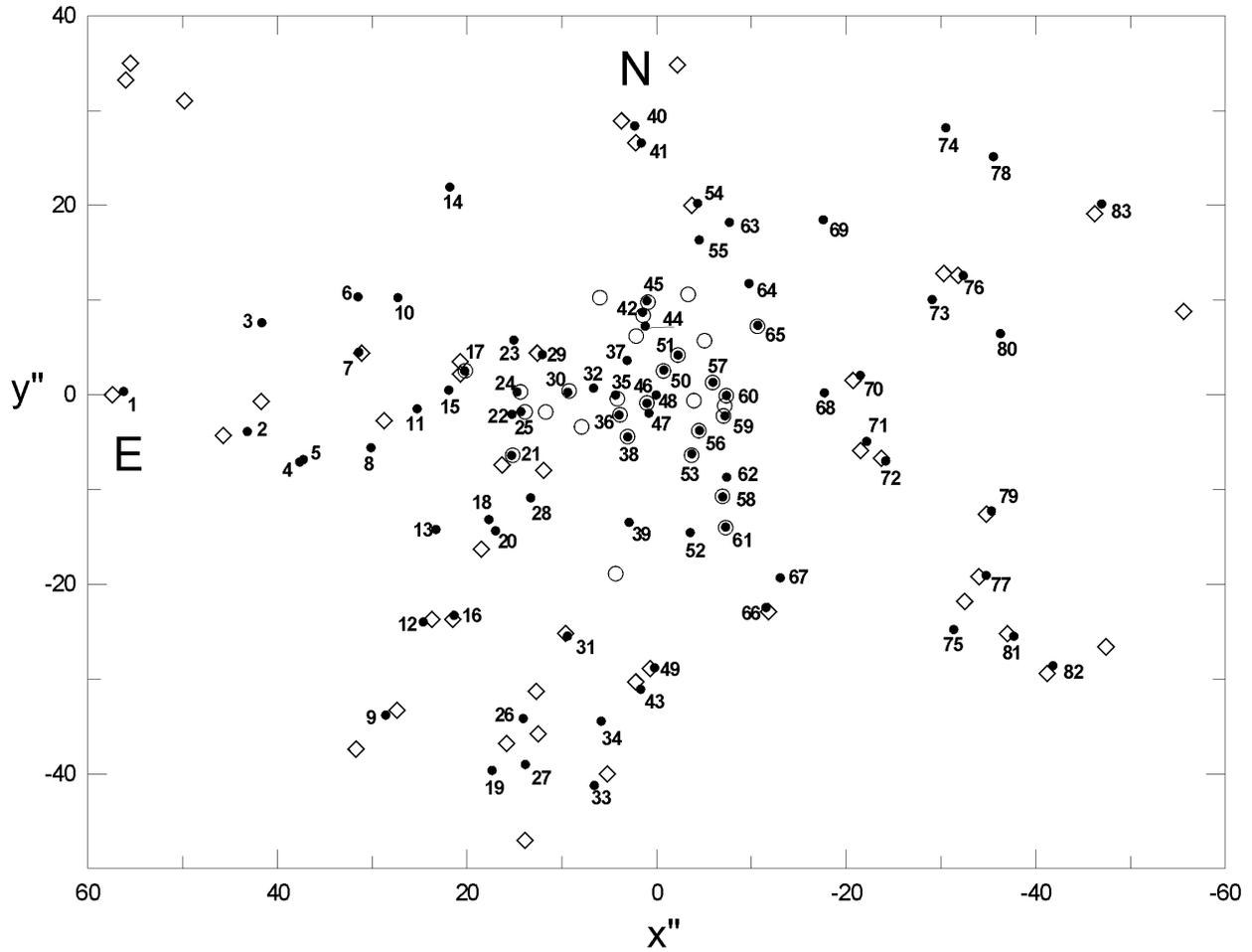}
\captionstyle{normal}
\caption{The identification of stars (1-83) from our list (denoted by dots
with the corresponding numbers) with known variables. The stars from the
catalog by Clement et al. (2001), except V128-155, are indicated by diamonds.
The stars from the list by Butler et al. (1998), 28 of which are variables
V128-155 and two most likely correspond to V83 and V85, are indicated by
circles. The differences $\Delta\alpha''$ and $\Delta\delta''$ relative to
the coordinates of the M15 center are used as $x''$ and $y''$ of the
rectangular coordinate system of the catalog by Clement et al. (2001).}
\end{figure}

\begin{figure}
\setcaptionmargin{5mm}
\onelinecaptionsfalse
\includegraphics[width=12.5cm]{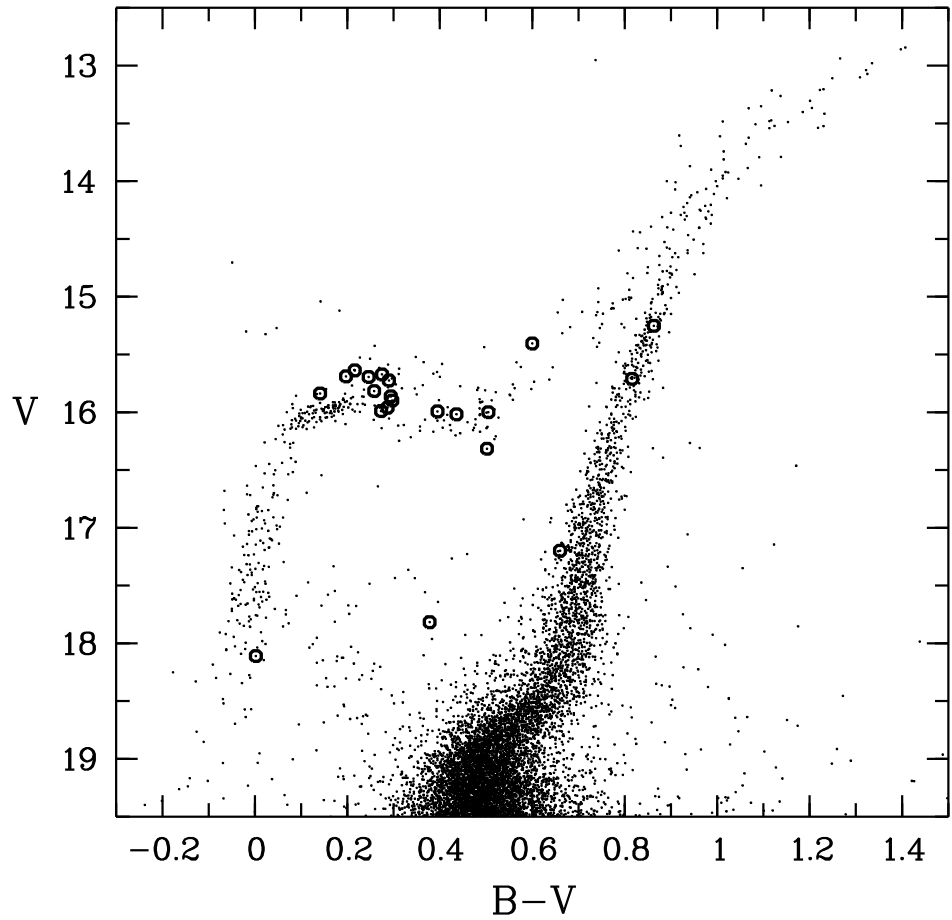}
\captionstyle{normal}
\caption{The $V-(B-V)$ diagram for the central region of M15 constructed by
using the catalog from van der Marel et al. (2002). The new variables
identifid with catalogued stars are indicated by circles.}
\end{figure}

\begin{figure}
\setcaptionmargin{5mm}
\onelinecaptionsfalse
\includegraphics[width=16.5cm]{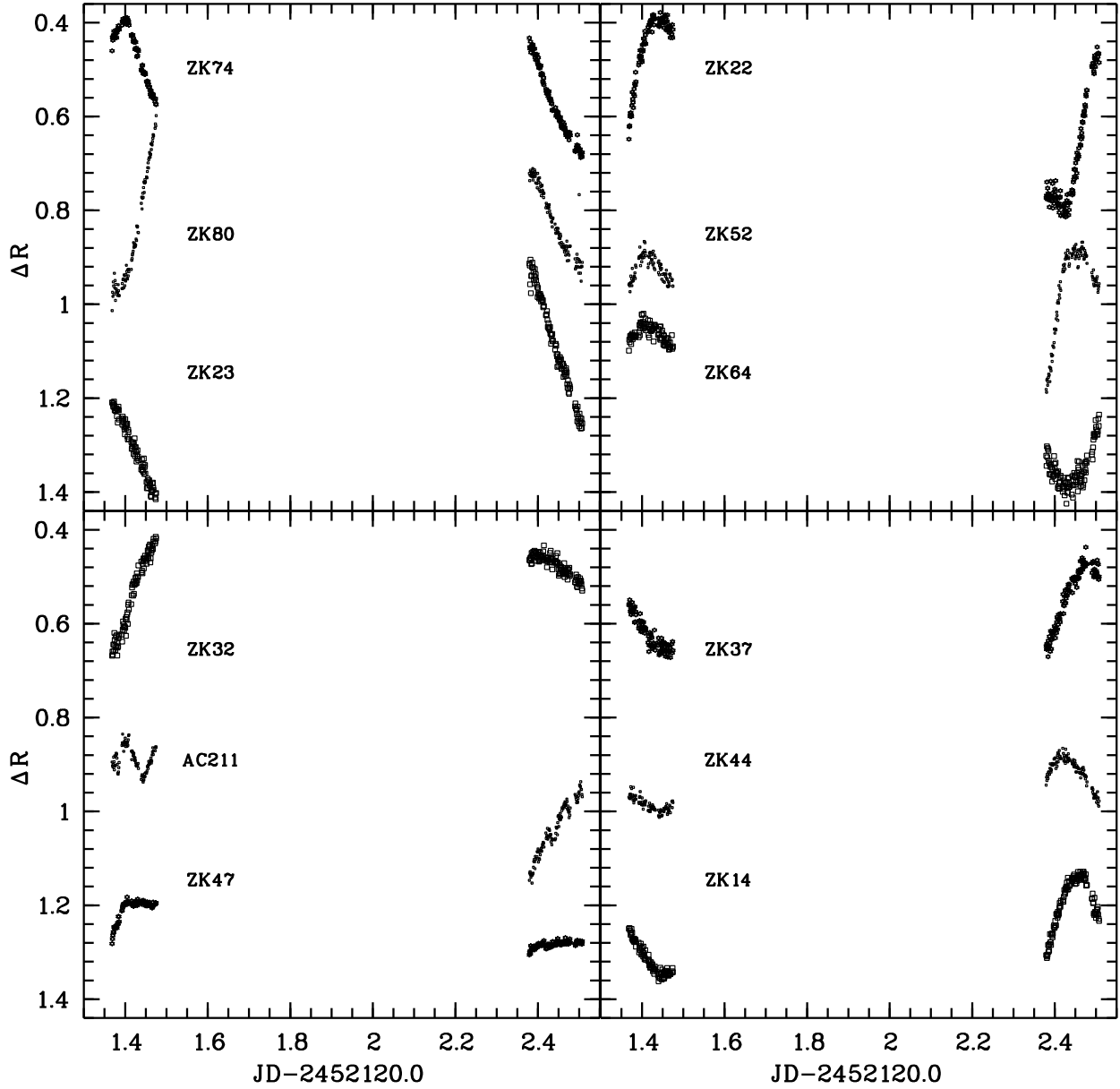}
\captionstyle{normal}
\caption{An illustration of the classification by types of new variables based
on their light curves. Three light curves for stars of each of the subtypes
(a) RR0, (b) RR1, and (d) RR2 are shown as an example. (c) The light curves
for the two stars whose variability type could not be determined; also shown
here is the light curve for the object AC 211. For convenience, the light
curves were arbitrarily displaced along the vertical axis (relative $R$
magnitudes).}
\end{figure}

\begin{figure}
\setcaptionmargin{5mm}
\onelinecaptionsfalse
\includegraphics[width=10.5cm]{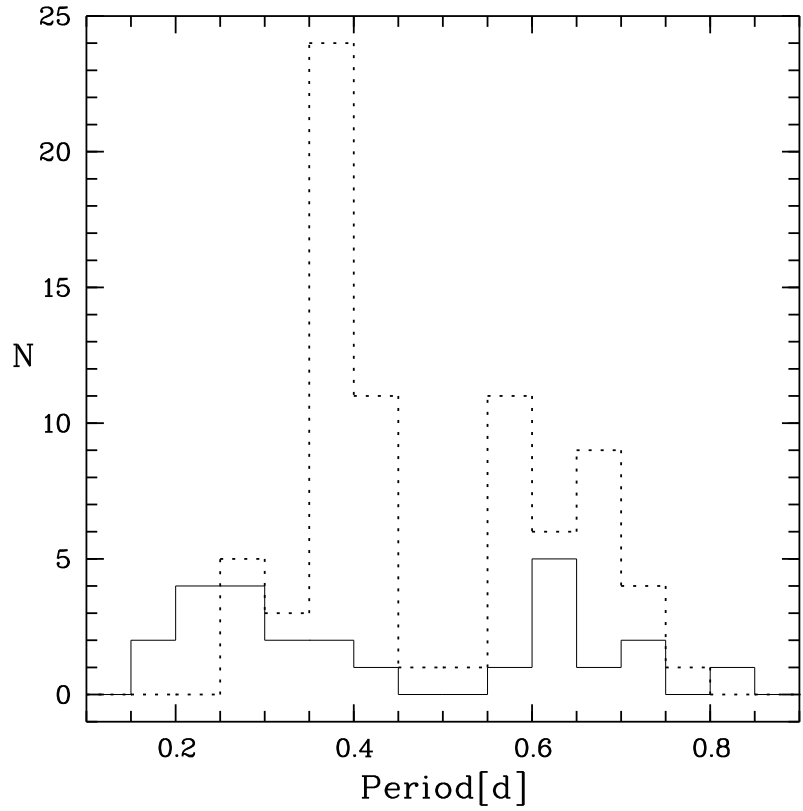}
\captionstyle{normal}
\caption{The period distributions (histograms) of RR Lyrae variables in the
globular cluster M15. The solid and dotted lines indicate the histograms
constructed for stars of the central cluster region ($r<35''$) and for the
remaining stars, respectively.}
\end{figure}

\begin{figure}
\setcaptionmargin{5mm}
\onelinecaptionsfalse
\includegraphics[width=7cm]{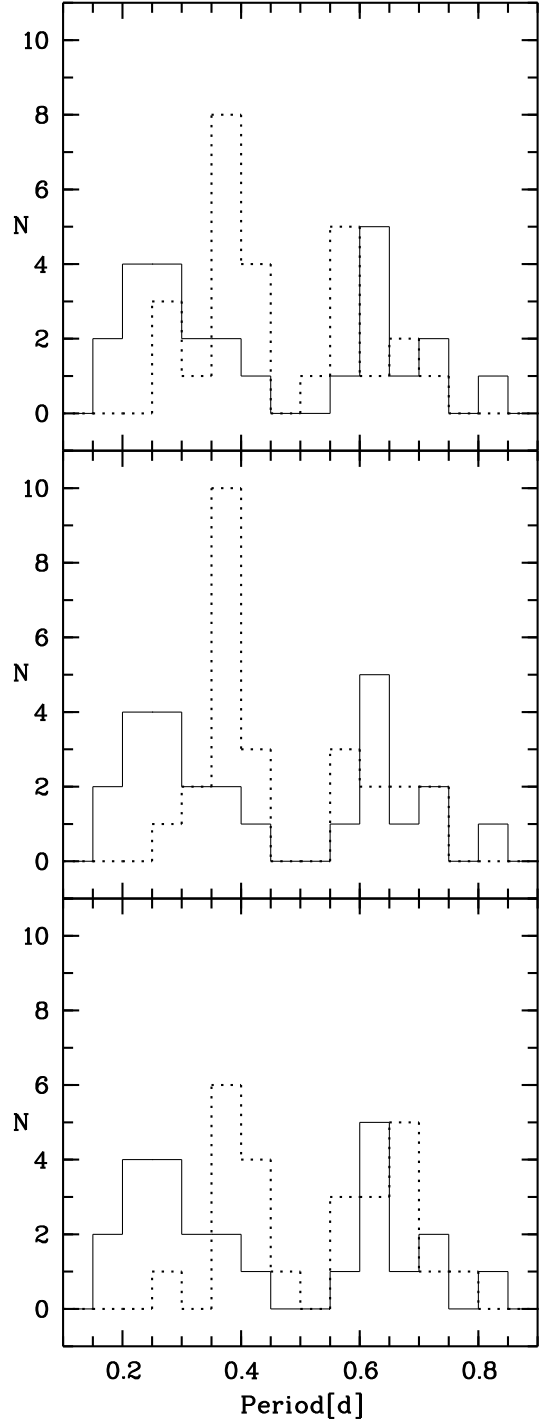}
\captionstyle{normal}
\caption{Comparison of the period distributions (histograms, dotted lines)
of RR Lyrae variables located in annular zones at difference distances
from the center of M15 ($r>175''$ (a), $90''<r<175''$ (b), and $35''<r<90''$
(c)) with a similar histogram (solid line) constructed for the same
variables in the central region ($r<35''$) of the cluster. Each zone,
except $r>175''$ (26 stars), contains 25 stars.}
\end{figure}

\begin{table}[p]
\setcaptionmargin{0mm}
\onelinecaptionsfalse
\captionstyle{flushleft}
\caption{Data on the 28 new variables in the central region of M15.
N are the numbers of the stars from our list classified as candidates
for new variables; $\Delta\alpha''$ and $\Delta\delta''$ are the
differences in equatorial coordinates relative to the coordinates of
the M15 center (AC 211, epoch 2000.0), which may be taken with a
sfficient accuracy as the rectangular ($x'', y''$) coordinates of the
system of the catalog by Clement et al. (2001); $\Delta R$ is the maximum
recorded $R$-band brightness variation, which is not an amplitude values
for some stars; the column "Type" gives a tentative determination of
the variability type.}
\bigskip
\begin{tabular}{|r|c|c|c|c|c|r|c|c|c|c|c|}
\hline
  N  & $\Delta\alpha''$ & $\Delta\delta''$ & $\Delta R$ & Type & N  &
$\Delta\alpha''$ & $\Delta\delta''$ & $\Delta R$ & Type  \\
\hline
ZK3  & 41.65 &   7.61 & 0\fm20 & RR1? & ZK39 &   2.91 & -13.46 & 0\fm28 &
RR1?   \\
ZK4  & 37.66 &  -7.09 & 0\fm45 & RR1? & ZK44 &   1.21 &   7.25 & 0\fm12 &
RR2?   \\
ZK5  & 37.26 &  -6.84 & 0\fm40 & RR1? & ZK47 &   0.81 &  -1.94 & 0\fm10 &
$?$  \\
ZK6  & 31.51 &  10.34 & 0\fm30 & RR0? & ZK52 &  -3.54 & -14.54 & 0\fm32 &
RR1?    \\
ZK10 & 27.30 &  10.25 & 0\fm12 & RR2? & ZK55 &  -4.49 &  16.33 & 0\fm25 &
RR2?    \\
ZK11 & 25.27 &  -1.47 & 0\fm22 & RR1?? & ZK62 &  -7.39 &  -8.70 & 0\fm25 &
SX Phe  \\
ZK13 & 23.29 & -14.21 & 0\fm30 & RR1? & ZK63 &  -7.68 &  18.19 & 0\fm28 &
RR1?   \\
ZK14 & 21.82 &  21.92 & 0\fm23 & RR2? & ZK64 &  -9.75 &  11.73 & 0\fm35 &
RR1?   \\
ZK18 & 17.70 & -13.16 & 0\fm52 & RR0?  & ZK67 & -13.04 & -19.32 & 0\fm30 &
RR1?   \\
ZK22 & 15.27 &  -2.06 & 0\fm42 & RR1?  & ZK68 & -17.70 &  0.20 & 0\fm20 &
SX Phe  \\
ZK23 & 15.07 &   5.78 & 0\fm52 & RR0?  & ZK69 & -17.58 &  18.47 & 0\fm30 &
RR1?  \\
ZK32 &   6.65 &  0.70 & 0\fm25 & $?$   & ZK74 & -30.51 &  28.17 & 0\fm30 &
RR0?  \\
ZK34 &   5.85 & -34.44 & 0\fm30 & RR2? & ZK78 & -35.53 &  25.13 & 0\fm32 &
RR1   \\
ZK37 &   3.12 &   3.63 & 0\fm20 & RR2? & ZK80 & -36.27 &  6.46 & 0\fm40 &
RR0?  \\

\hline

\end{tabular}
\end{table}

\begin{table}[p]
\setcaptionmargin{0mm}
\onelinecaptionsfalse
\captionstyle{flushleft}
\caption{Corrected coordinates of the variable stars V128-155 from the
catalog by Clement et al. (2001). N are the numbers of the catalogued
stars; $\Delta\alpha''$ and $\Delta\delta''$ are the same as in Table 1.}
\bigskip
\begin{tabular}{|r|c|c|r|c|c|r|c|c|}
\hline
 N   & $\Delta\alpha''$ & $\Delta\delta''$ &  N  & $\Delta\alpha''$ &
$\Delta\delta''$ &  N  & $\Delta\alpha''$ & $\Delta\delta''$   \\
\hline
128 &  1.03 &  -0.86 & 138 &  3.07 &  -4.12 & 148 &  -3.93 &  -0.62  \\
129 & -7.28 & -14.02 & 139 &  1.43 &   8.38 & 149 &   2.16 &   6.20  \\
130 & -6.95 & -10.71 & 140 &  4.15 &  -0.42 & 150 &   5.98 &  10.25  \\
131 & -7.05 &  -2.23 & 141 &  9.23 &   0.41 & 151 &  -7.18 &  -1.15  \\
132 & -7.35 &  -0.09 & 142 &  3.86 &  -2.11 & 152 &   0.92 &   9.77  \\
133 & -5.90 &   1.30 & 143 &  4.32 & -18.88 & 153 &  -5.05 &   5.71  \\
134 & -4.44 &  -3.79 & 144 & 13.89 &  -1.81 & 154 &  -3.32 &  10.60  \\
135 & -3.68 &  -6.38 & 145 & 14.38 &   0.31 & 155 & -10.64 &   7.23  \\
136 & -2.24 &   4.20 & 146 & 11.69 &  -1.81 &     &        &         \\
137 & -0.71 &   2.52 & 147 &  7.91 &  -3.38 &     &        &         \\
\hline

\end{tabular}
\end{table}

\end{document}